\documentclass[twocolumn]{article}
\usepackage{amsmath,amssymb,subfigure}
\usepackage[dvips]{graphicx}
\usepackage[english]{babel}

\long\def\comment#1{}
 
\makeatletter
\def\@normalsize{\@setsize\normalsize{10pt}\xpt\@xpt
\abovedisplayskip 10pt plus2pt minus5pt\belowdisplayskip 
\abovedisplayskip \abovedisplayshortskip \z@ 
plus3pt\belowdisplayshortskip 6pt plus3pt 
minus3pt\let\@listi\@listI}
 
\def\subsize{\@setsize\subsize{12pt}\xipt\@xipt}
\def\section{\@startsection {section}{1}{\z@}{1.0ex plus
1ex minus .2ex}{.2ex plus .2ex}{\large\bf}}
\def\subsection{\@startsection 
   {subsection}{2}{\z@}{.2ex plus 1ex} {.2ex plus .2ex}{\subsize\bf}}
\makeatother

\begin{document}
 
\title{\Large {\bf Stability of Thermohaline Circulation with respect to fresh water release }}

\author{{Ajay Patwardhan, Vivek Tewary}\\\small{St. Xavier's College, Mumbai}}

\maketitle
\thispagestyle{empty}

\subsection*{\centering Abstract}
{\em 
The relatively warm climate found in the North-Western
Europe is due to the Gulf Stream that circulates warm
saline water from southern latitudes to Europe. In North
Atlantic Ocean the stream gives out a large amount of heat,
cools down and sinks to the bottom to complete the
Thermo-Haline Circulation. There is considerable debate
on the stability of the stream to inputs of fresh water from
the melting ice in Greenland. The circulation, being
switched off, will have massive impact on the climate of
Europe. Intergovernmental Panel on Climate Change
(IPCC) has warned of this danger in its recent report. Our
aim is to model the Thermo-Haline Circulation at the point
where it sinks in the North-Atlantic. We create a two-
dimensional discrete map modeling the salinity gradient and
vertical velocity of the stream. We look for how a
perturbation in the form of fresh water release can
destabilize the circulation by pushing the velocity below a
certain threshold.
}
\section{Introduction}
\label{Introduction}

In the ocean, water motion is not only generated by wind-forcing. The horizontal
temperature and salinity differences caused by climatic influences at the sea surface, cause
density differences which initiate circulation. Such circulation requires that expansion
should take place at a higher pressure than contraction does. In terms of ocean and water
temperature, this means that the heat source must lie at a lower level than the cold
source. In the ocean, the heat and cold sources are located at the same level, at the sea
surface. Water close to a heat source attains a lower density than water close to a cold
surface. It becomes lighter and spreads at the surface in the direction of the cold source.
For continuity, water below the heat source will ascend and water below the cold source
will descend and while spreading below, it will become warmer through heat conduction
and mixing. In the layers down below, cold water moves from the higher latitudes to the
lower ones and in the upper layer, warm water flows in the opposite direction. This is not
the entire story. At the sea surface globally, there are regions rich in salinity as well as
those deficient of salinity. The former has more evaporation and ice formation (a cold
source) and the latter has more precipitation, continental run-offs and ice-melting (a heat
source). This causes a haline circulation.
In places, the Thermal and the Haline circulations act in the same direction, reinforcing
each other to form what is called the Global Thermohaline Circulation (THC). It is often
referred to as the Great Conveyer Belt.In the North Atlantic, it manifests
itself as the meridional overturning
current (MOC).                      This current,
through the Gulf Stream and the
North Atlantic Current, transports
large quantities of warm water to
the northern latitudes. This has a
strong     effect    on    climatic
conditions. Compared to the
corresponding parts of North
America, Northern Europe has a
much milder climate due to this
heat transport. It has been
proposed by the Intercontinental
Panel on Climate Change (IPCC)
that the challenge of global
warming is real and it is man-
made. There are concerns~\cite{pa} that
global warming can affect the
stability of the MOC. During the
last deglaciation (whose remnants
are the Great Lakes in the North
America), a large amount of melt
water entered the North Atlantic
causing a shallower THC cell. North Atlantic’s ice cover has significantly changed over the
past 20,000 years resulting in drastic climatic changes. However the ice cover over
Greenland is much more massive than the polar sea ice cover. Melt water from Greenland
could trigger a switchover of the MOC. 

\section{The Model}
\label{The Model}
The aim here is to create a simple 2-D discrete
model of MOC. The amount of scaled fresh water intake is taken as a parameter. The
impact of its variation is studied. Our starting point is a modification of an MOC model by Timmerman, Lohmann and Monahan~\cite{pal,pala,palan}.
\begin{align}
\dot{x}=x(1-x)+\mu+\frac{y}{h} 
\\
\dot{y}=\frac{h}{T^2}x-g+\frac{\alpha}{T}y
\end{align}
Discretization yields
\begin{align}
x_\text{n+1}=ax_\text{n}(1-x_\text{n})+\mu+\frac{y_\text{n}}{h}\\
y_\text{n+1}=\frac{h}{T^2}x_\text{n}-g+\frac{\alpha}{T}y_\text{n}
\end{align}
Now we shall fix some values for the parameters of the map. As the depth to which the
meridional overturning descends is approximately 3 km, we shall take h = 3000. We shall
take the proportionality constant '$\alpha$' as 1. T, we take as 20. g, we keep as the acceleration
due to gravity on Earth surface, 9.8. With these values, the map much simplifies to
\begin{align}
x_\text{n+1}=ax_\text{n}(1-x_\text{n})+\mu+\frac{y_\text{n}}{3000}\\
y_\text{n+1}=7.5x_\text{n}-9.8+0.05y_\text{n}
\end{align}
\subsection{Low Fresh Water Intake}
\label{Low Fresh Water Intake}
The variation of the map with ‘a’ is not of much interest to us.$\mu$ is a scaled variable and
we shall study its variation in the range $(0,1)$. We begin with $\mu=0.01$ and $a=1.1$.
The reason for the choice of '$\mu$' is that we are talking of the North Atlantic, where
freezing rather than melting predominates, and physically that means a small, if any, fresh
water intake for THC.The fixed points of the system are \begin{align*}
\bar{x_1}=-0.0435779,\bar{y_1}=-10.6598\\
\bar{x_2}=0.136879,\bar{y_2}=-9.23516
\end{align*}
The first fixed point is a saddle point, however, the second fixed point is stable.For the unstable fixed point, if we start with a smaller value than that fixed point, the
orbit diverges to negative infinity for both variables. But starting with a value more than
this fixed point, the orbit converges to the stable fixed point(See Figure 1).\\
\begin{figure*}
\centerline{
\mbox{\includegraphics[width=3.00in]{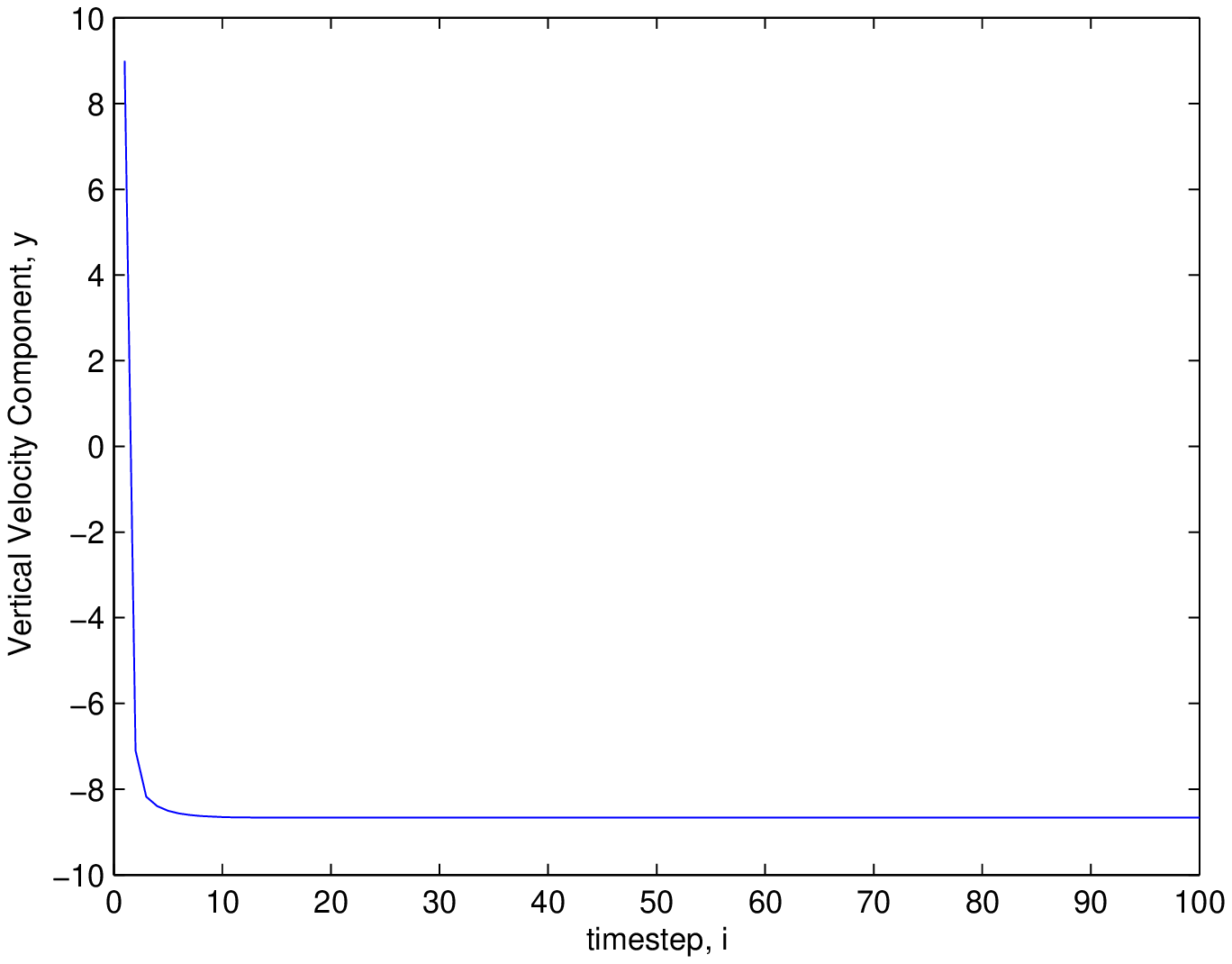}}
\mbox{\includegraphics[width=3.00in]{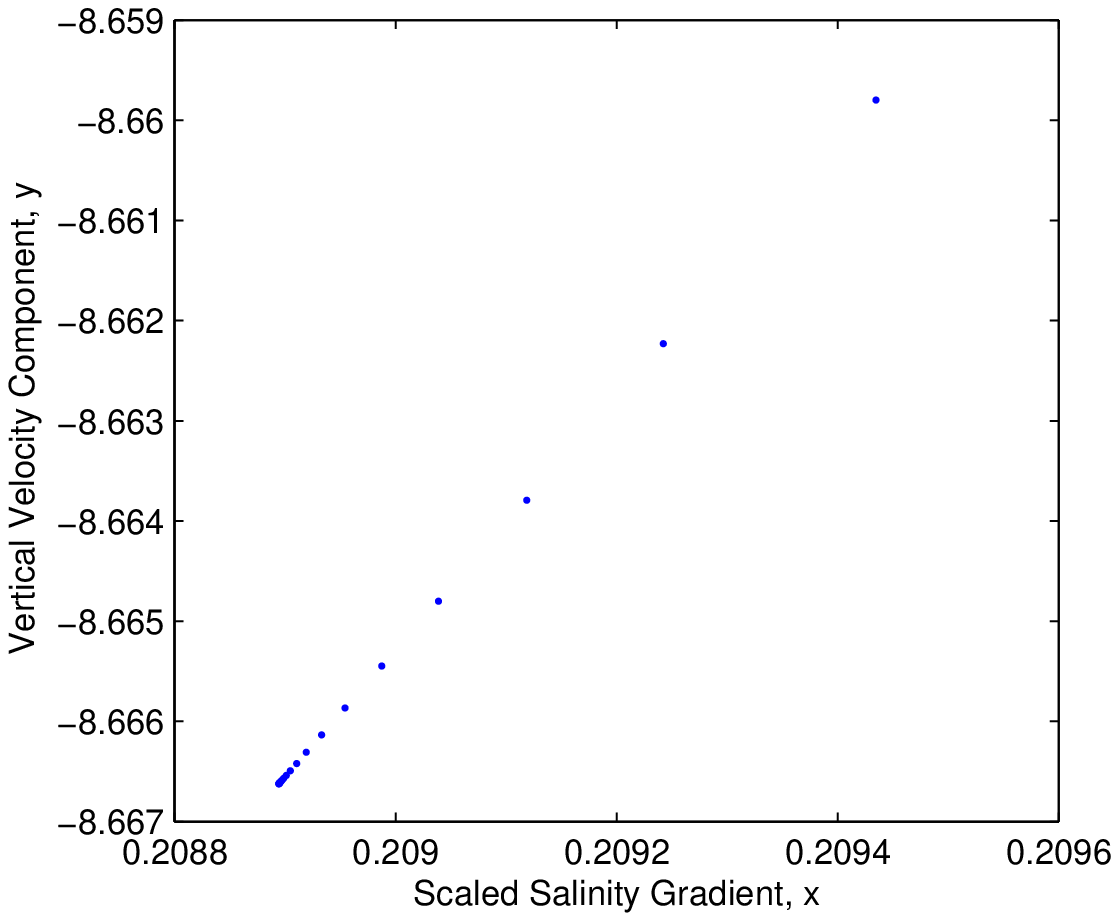}}
}
\caption{Time Series and Phase Plot showing convergence for low fresh water intake}
\label{overView}
\end{figure*}
\subsection{High Fresh Water Intake}
\label{High Fresh Water Intake}
Now our intent is to find out what happens to the system if suddenly a large amount of
melt water finds its way into the MOC. This would correspond to a large value of $\mu$. $\mu=0.8$ is chosen. This corresponds to a sudden inflow of melt water from the Greenland ice cover. This is a low probability situation, but, in any case accentuated by the effects of global warming.The fixed points of the system are found to be
\begin{align*}
\bar{x_1}=-0.805595,\bar{y_1}=-16.6758\\
\bar{x_2}=0.898897,\bar{y_2}=-3.21924
\end{align*}
This time also, the second point is a stable fixed point, however, the first fixed point is a
saddle and therefore unstable. As can be seen, the stable solution is one with a high
salinity gradient, which is to be expected on addition of fresh water to the system and a
lower velocity, signifying the weakening of the MOC. Now suppose, we were to begin
from the previous stable fixed point, it would be of interest to see how the system
evolves.This time the system settles down into the new fixed point in a matter of fifty time steps(See Figure 2).\\
\begin{figure*}
\centerline{
\mbox{\includegraphics[width=3.00in]{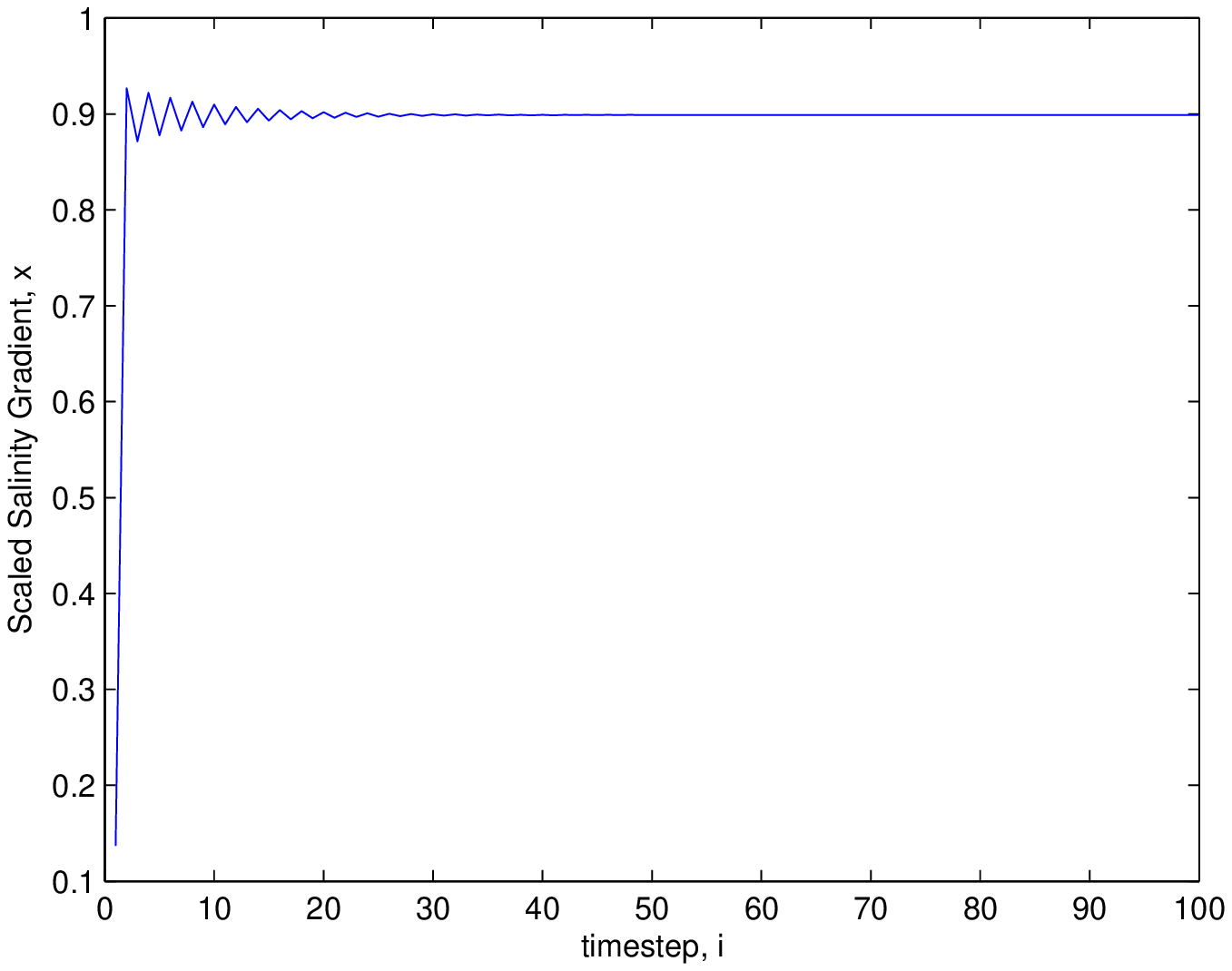}}
\mbox{\includegraphics[width=3.00in]{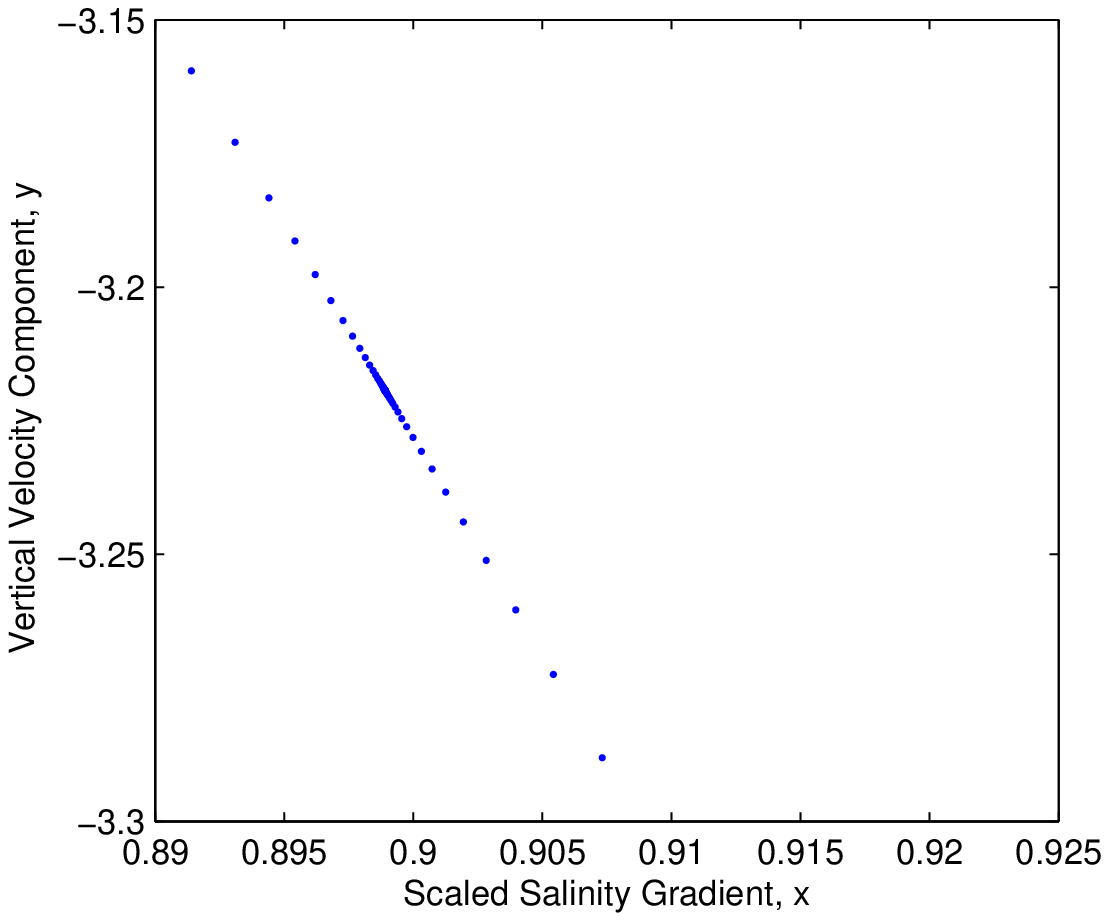}}
}
\caption{Time Series and Phase Plot showing convergence for high fresh water intake}
\label{overViewe}
\end{figure*}
\subsection{Bifurcation Plot}
\label{Bifurcation Plot}
We are interested in how the stable fixed points of the map vary with change in $\mu$. This
can be visualised by the Bifurcation Diagram for the range of (0, 1).The bifurcation diagram shows a smooth slowing down of the stream velocity with respect to $\mu$ . Close to $\mu\sim0.9$ we can see the first bifurcation of the map(See Figure 3).It is seen that the bifurcation parameter $\mu_B$ remains almost fixed at $\sim0.9$ with respect to variation in the parameter 'T'. Thus, the system is robust with repsect to changes in the parameter 'T'.
\begin{figure*}
\centering
\subfigure[T=20] 
{
    \label{T=20}
    \includegraphics[width=6cm]{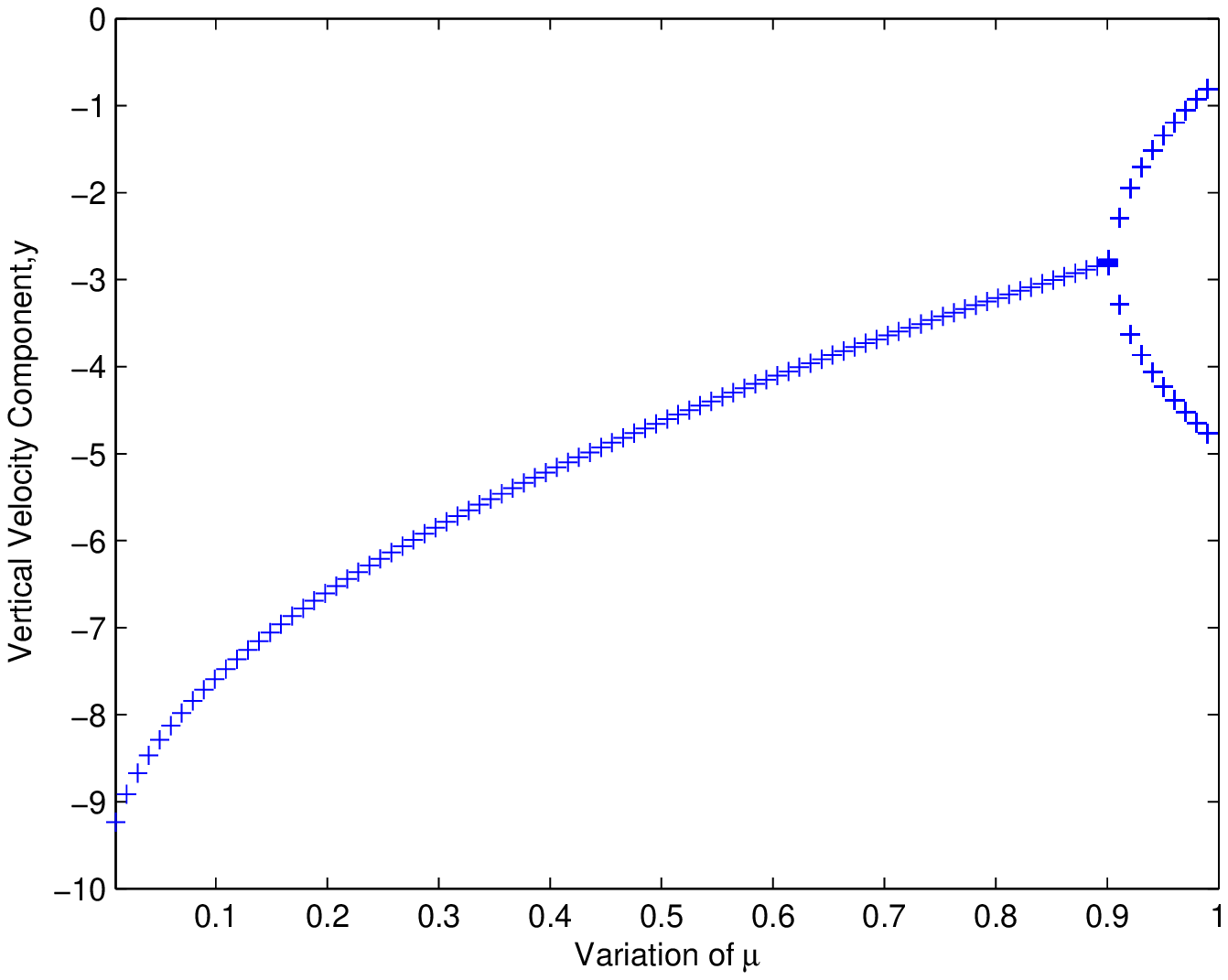}
}
\hspace{1cm}
\subfigure[T=100]
{
    \label{T=100}
    \includegraphics[width=6cm]{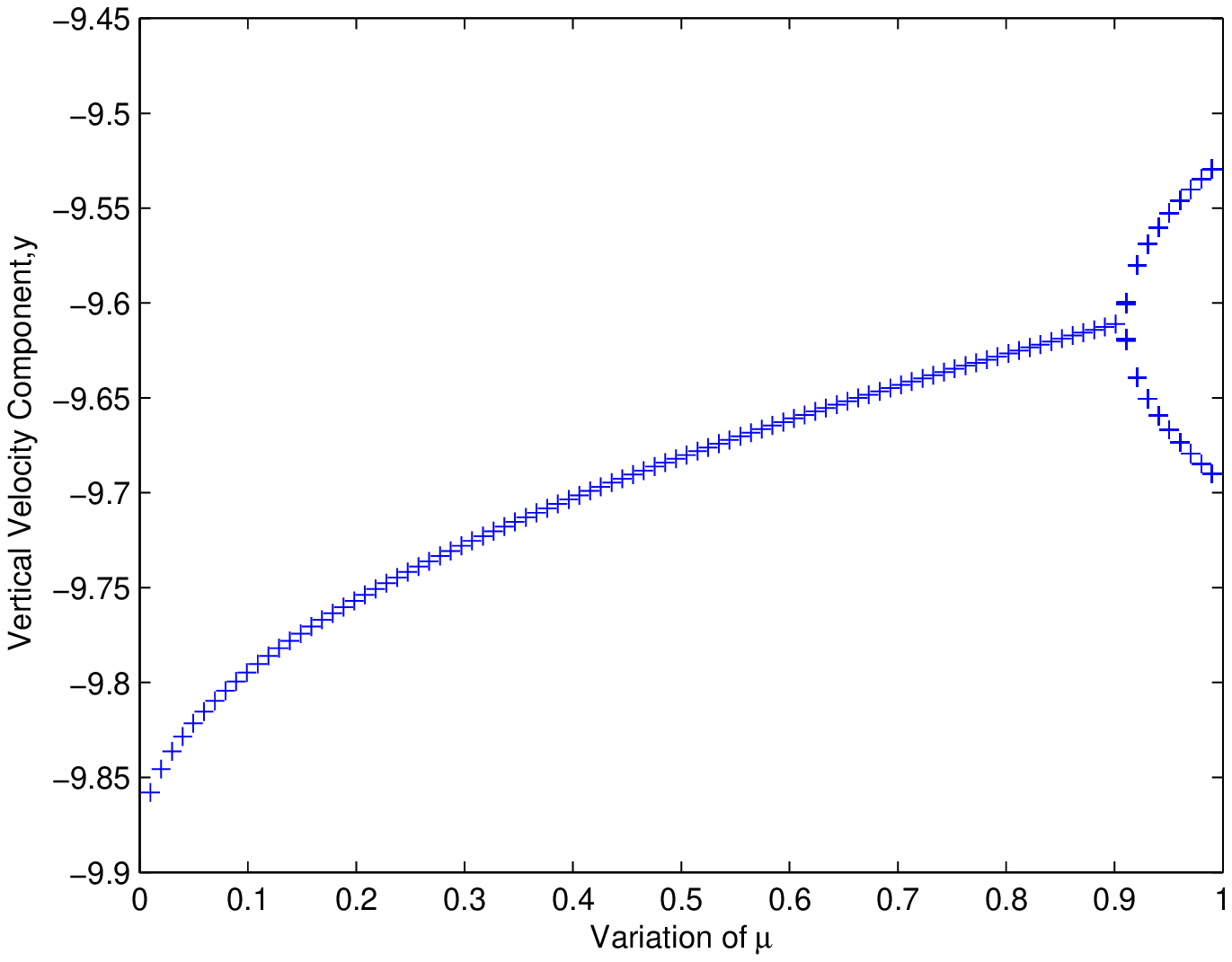}
}
\hspace{1cm}
\subfigure[T=500]
{
    \label{T=500}
    \includegraphics[width=6cm]{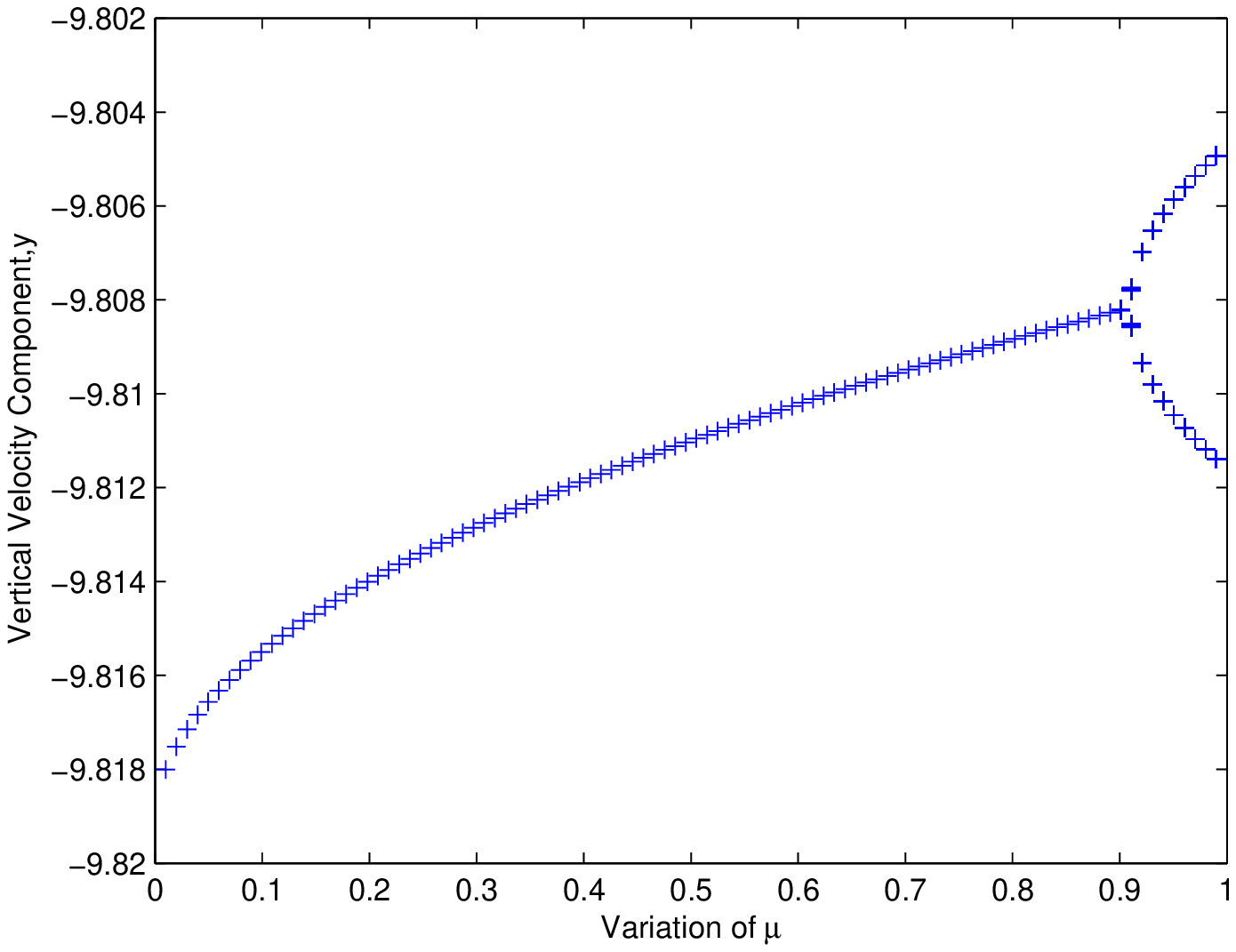}
}
\hspace{1cm}
\subfigure[T=1000]
{
    \label{T=1000}
    \includegraphics[width=6cm]{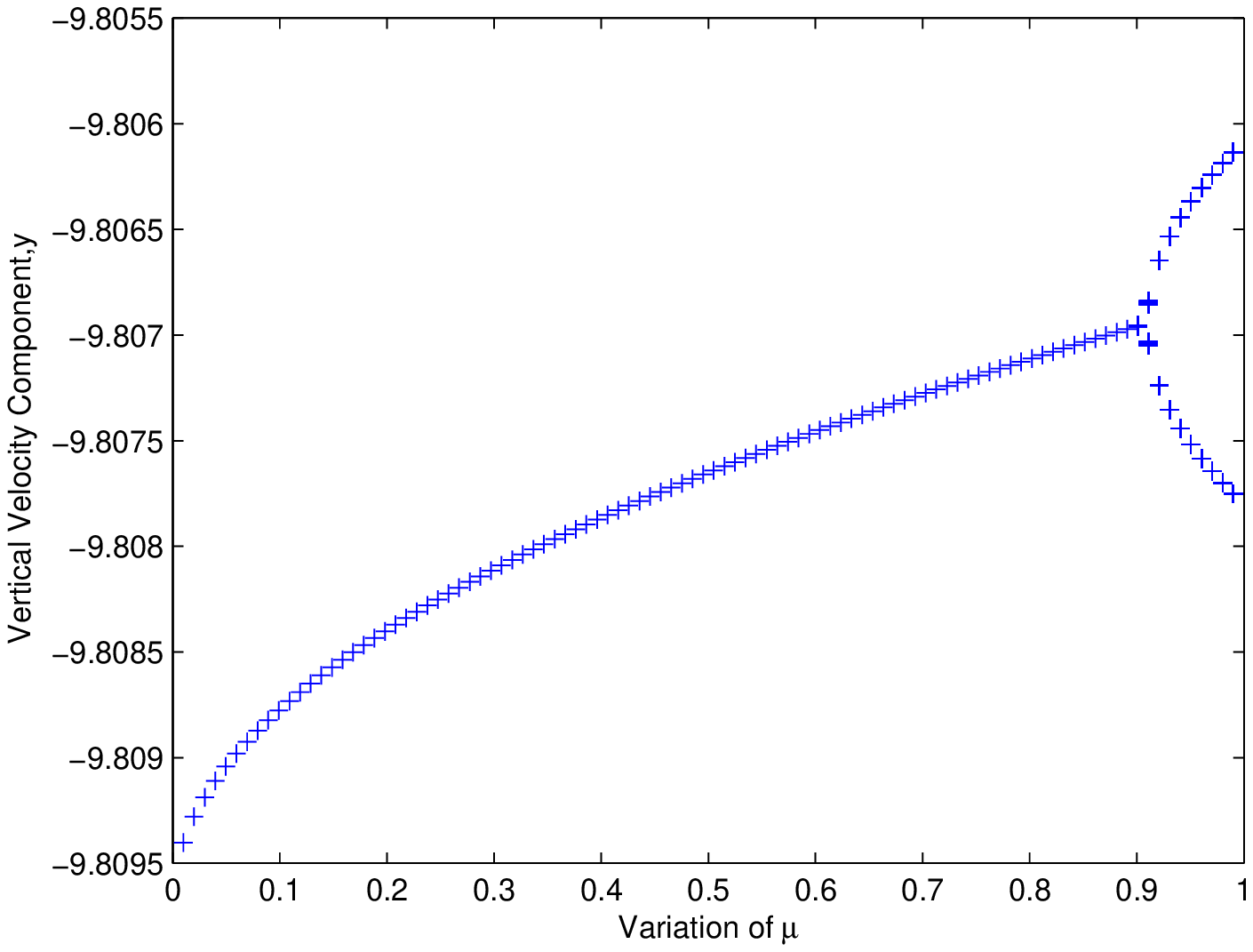}
}
\caption{Bifurcation Plot showing variation of vertical velocity with change in $\mu$ for different values of parameter 'T'}
\label{overViewer}
\end{figure*}
\section{Conclusion}
\label{Conclusion}
The interpretation that we make of this is that a gradual change in µ would cause the
stream to slow down only gradually and only a sudden deglaciation leading to release of a
massive amount of fresh water can cause the Gulf Stream to slow down. Therefore, it can
be said that this is a low probability but disastrous event. What we would like to further
do with the map is to standardise the values of the variables using real values and to see
how the discrete time step translates into physical time because this would give us some
idea of the time in which the system can get destabilised.

\end{document}